\documentclass[onecolumn,floatfix,superscriptaddress,a4paper,
               showpacs,showkeys,nofootinbib,preprint]{revtex4}
\usepackage{epsfig}
\usepackage{amssymb,latexsym,amsmath}
\newcommand{\eq}[1]{\begin{align} #1 \end{align}}

\begin{document}

\title{%Second Class of
Strongly Intensive Quantities}

 \author{M. I. Gorenstein}
 \affiliation{Bogolyubov Institute for Theoretical Physics, Kiev, Ukraine}
 \affiliation{Frankfurt Institute for Advanced Studies, Frankfurt, Germany}

 \author{M. Ga\'zdzicki}
 \affiliation{Institut f\"ur Kernphysik, University of Frankfurt, Frankfurt, Germany}
 \affiliation{Jan Kochanowski University, Kielce, Poland}

\begin{abstract}

Analysis of fluctuations of hadron production properties in
collisions of relativistic particles profits from use of
measurable intensive quantities which are independent of system
size variations. The first family of such quantities was proposed
already in 1992. The second is introduced in this paper. We also
present a proof of independence of volume fluctuations for
quantities from both families within the framework of the grand
canonical ensemble. These quantities are referred to as strongly intensive
ones.
Influence of conservation laws and resonance decays is also discussed. 

\end{abstract}



\maketitle

\section{Introduction}
An intensive quantity is a physical quantity which does not depend
on the system volume.  By contrast, an extensive quantity is
proportional to the system volume. Clearly, the ratio of two
extensive quantities is an intensive one. 
For example the number of particles, $N$,  in the
relativistic gas fluctuates around its mean value, $\langle N \rangle$. 
Within the grand canonical ensemble $\langle N \rangle$ is an
extensive quantity, whereas the ratio of mean multiplicities of
two different particle types is an intensive one. Particle
number fluctuations are quantified by the variance, $\langle N^2
\rangle - \langle N \rangle^2$, which is an extensive quantity,
while the scaled variance, \mbox{$[\langle N^2 \rangle - \langle N
\rangle^2] / \langle N \rangle$}, is an intensive one.

Statistical models are surprisingly successful in modeling
multi-particle production in high energy
interactions~\cite{florkowski}. They are used to describe
properties of  strongly interacting matter created in
nucleus-nucleus collisions in terms of intensive quantities. 
In particular, an equation of state is usually given as a function
relating pressure, temperature and baryonic chemical potential. On
the other hand, in  high energy collisions the volume of produced
matter cannot be kept fixed. For instance, nucleus-nucleus
collisions with different centralities may produce a statistical
system with the same local properties (e.g., the same temperature
and baryonic chemical potential) but with the system volume
changing significantly from interaction to interaction. Thus, an
important question is whether it is possible to measure the
properties of the system without knowing its volume fluctuations,
or equivalently, whether there are measurable quantities which are
independent of volume fluctuations.

Within the grand canonical ensemble
the answer is yes, and the quantities with the required properties
will be referred to as strongly intensive ones. Ratios of mean
particle multiplicities are  strongly intensive quantities. In
general, this is the case for all ratios of any two extensive
quantities which correspond to the first moments of fluctuating
variables. They are intensive and strongly intensive quantities.
The situation is, however, more complicated for the measures of
fluctuations which include the second moments of fluctuating
variables. For example, as will be shown below, the scaled
variance of a particle number distribution is an intensive quantity, but
not a strongly intensive one.

In this paper we  show that there are two families of strongly
intensive quantities which characterize the second moments of
random extensive variables used to study fluctuations and correlations
in a physical system. While the first family was introduced
already in 1992~\cite{GM}, the second one is proposed in this paper.

The paper is organized as follows. In Sec.~II the two families of
strongly intensive quantities are introduced. For simplicity,
this is done within the model of independent particle
sources. The relation of the strongly intensive quatities
to previously used fluctuation measures is
discussed in Sec.~III. The  proof that the quantities are in fact
strongly intensive, i.e., strictly independent of volume and
volume fluctuations within the grand canonical ensemble is given
in Sec.~IV. Finally, their properties within the canonical and
micro-canonical ensembles are discussed in Sec.~V. Summary given in
Sec.~VI closes the paper.

\section{Two families of strongly intensive quantities}
Let us start from the model of independent sources for
multi-particle production in which the number of sources, $N_s$,
changes from event to event. In this model, extensive quantities
(e.g., mean number of particles, mean transverse energy) will be
considered as those which are proportional to $N_s$. Two
fluctuating extensive variables $A$ and $B$ can be expressed as:
\eq{\label{ind-sour}
A~=~a_1~+a_2~+\ldots~+ a_{N_s}~,~~~~~B~=~b_1~+b_2~+\ldots~+
b_{N_s}~,
}
where  $a_k$ and $b_k$ denote the contributions from the $k$-th
source. One finds for event averages:
 \eq{\label{A}
 &\langle A \rangle~=~\langle a \rangle ~\langle N_s\rangle~,~~~~
 \langle A^2 \rangle~=~\langle a^2 \rangle ~\langle N_s \rangle ~+~
 \langle a \rangle ^2
 ~\left[\langle N_s^2\rangle ~-~\langle N_s\rangle\right]~,\\
&\langle B \rangle~=~\langle b \rangle ~\langle N_s\rangle~,~~~~
 \langle B^2 \rangle~=~\langle b^2 \rangle ~\langle N_s \rangle ~+~
 \langle b \rangle ^2
 ~\left[\langle N_s^2\rangle ~-~\langle N_s\rangle\right]~,\label{B} \\
 &\langle AB \rangle~=~
\langle a~b \rangle ~\langle N_s \rangle ~+~
 \langle a \rangle \langle b \rangle
 ~\left[\langle N_s^2\rangle ~-~\langle
 N_s\rangle\right]~,\label{AB}
}
where  $\langle a\rangle$, $\langle b\rangle$ and $\langle
a^2\rangle$, $\langle b^2\rangle$, $\langle ab\rangle$  are the
first and second moments of the distribution $P^*(a,b)$ for a
single source. These quantities are independent of $N_s$ and play
the role of intensive quantities in the model of independent
sources. The distribution $P^*(a,b)$ is assumed to be the same for
all sources, i.e., they are statistically identical. The
probability distribution $P_s(N_s)$ of the source number is needed
to calculate $\langle N_s\rangle $ and $\langle N_s^2\rangle$ and,
in general, it is unknown.
Using Eq.~(\ref{A}) the scaled variance $\omega_A$ which describes
the event-by-event fluctuations of the extensive variable $A$ can
be presented as:
 \eq{\label{omega-A}
  \omega_A~\equiv~ \frac{\langle A^2\rangle
~-~\langle A \rangle ^2}{\langle A\rangle}~=~\frac{\langle
a^2\rangle~-~\langle a\rangle^2}{\langle a\rangle}~+~\langle a
\rangle ~\frac{\langle N_s^2\rangle ~-~\langle
N_s\rangle^2}{\langle N_s\rangle} ~\equiv~\omega_a^*~+~\langle a
\rangle~ \omega_s~,
}
where $\omega_a^*$ is the scaled variance of the quantity $A$ for each
source. A similar expression follows from Eq.~(\ref{B}) for the
scaled variance $\omega_B$. The scaled variances $\omega_A$ and
$\omega_B$ are independent of the average number of sources
$\langle N_s\rangle$. Thus, $\omega_A$ and $\omega_B$ are
intensive quantities. However, they depend on the fluctuations of
the number of sources  via $\omega_s$ and, therefore, they are not
strongly intensive quantities.

From Eq.~(\ref{A}) follows that a knowledge of $\langle A\rangle$
and $\langle A^2\rangle$ is not sufficient to derive any strongly
intensive quantity. This is, however, possible when two extensive
random variables, $A$ and $B$ are considered.  In order to characterize
fluctuations, one may then construct special combinations of the
second moments in which the terms proportional to $\langle
N^2_s\rangle$ in the r.h.s. of Eqs.~(\ref{A}-\ref{AB}) are not
present. Clearly, only two linearly independent combinations of
this type result from three equations~(\ref{A}-\ref{AB}). Note
that in order to remove the dependence on $\langle N_s\rangle$,
strongly intensive quantities should be in a form of reducible
fractions. The following combinations seem the most convenient:
\eq{\label{Psi}
\Sigma^{AB}~&=~\langle C\rangle^{-1}~\left[~\langle
B\rangle~\omega_A ~+~\langle A\rangle~ \omega_B ~-~2\left(~\langle
A B \rangle -\langle A\rangle\langle
B\rangle~\right)~\right]~,\\
\Delta^{AB}~&=~{\langle C\rangle}^{-1}~ \left[ ~\langle B\rangle~
\omega_A ~-~\langle A\rangle ~\omega_B ~\right]~, \label{G}
}
where $\langle C\rangle$ is the average of any extensive quantity,
e.g., $\langle A\rangle$ or $\langle B\rangle$. Straightforward
calculation of (\ref{Psi}) and (\ref{G}) using
Eqs.~(\ref{A}-\ref{omega-A}) gives:
 \eq{\label{Psi1}
\Sigma^{AB}~&=~\langle c\rangle^{-1}~\left[~\langle
b\rangle~\omega_a^* ~+~\langle a\rangle~ \omega_b^*
~-~2\left(~\langle ab \rangle -\langle a\rangle\langle
b\rangle~\right)~\right]~,\\
\Delta^{AB}~&=~{\langle c\rangle}^{-1}~ \left[ ~\langle b\rangle~
\omega_a^* ~-~\langle a\rangle ~\omega_b^* ~\right]~. \label{G1}
}
Thus, $\Sigma^{AB}$ and $\Delta^{AB}$ defined by
Eqs.~(\ref{Psi}-\ref{G}) depend on $\omega_a^*$ and $\omega_b^*$,
but they are independent of the average number of sources $\langle
N_s \rangle$ and  its fluctuations $\omega_s$. They are, in fact,
strongly intensive measures which quantify fluctuations of any two
extensive random variables $A$ and $B$. This will be proved in Sec.~IV
within the grand canonical ensemble for a case of particle
multiplicities.

There is an important  difference between the $\Sigma^{AB}$ and
$\Delta^{AB}$ quantities. Namely, in order to calculate
$\Delta^{AB}$ one  needs to measure only the first two moments:
$\langle A\rangle$, $\langle B\rangle$ and $\langle A^2\rangle$,
$\langle B^2\rangle$. This can be done by  independent
measurements of the distributions $P_A(A)$ and $P_B(B)$. Quantity
$\Sigma^{AB}$ includes the correlations term, $\langle AB \rangle
-\langle A \rangle \langle B \rangle$, and thus it requires, in
addition,  simultaneous measurements of $A$ and $B$ in order to
obtain the joint probability distribution $P_{AB}(A,B)$. The quantities
$\Sigma^{AB}$ and $\Delta^{AB}$ also have different symmetry
properties: $\Sigma^{AB}=\Sigma^{BA}$ and $\Delta^{AB}= -~
\Delta^{BA}$. We call all strongly intensive quantities which
include the correlation term the $\Sigma$ family, those which
include only variances the $\Delta$ family.
\section{Relation to other fluctuation measures}
The well-known fluctuation measure $\Phi$ was introduced in
1992~\cite{GM} for a study of transverse momentum fluctuations and
it belongs to the $\Sigma$ family. In the general case~\cite{liu},
when $A\equiv X=\sum_{i=1}^N x_i$ represents any motional
extensive variable as a sum of single particle variables, and
$B\equiv N$ is the particle multiplicity, one gets:
\begin{equation}\label{phi_x}
\Phi_x~=~\left[ \frac{\langle X \rangle} {\langle N\rangle
}~\Sigma^{XN}\right]^{1/2}~-~\left[\overline{x^2}~-~\overline{x}^2\right]^{1/2}~,
\end{equation}
where $\Sigma^{XN}$ is given by Eq.~(\ref{Psi}) with $C\equiv N$,
and $\overline{x^2}$, $\overline{x}^2$ correspond to single-particle
inclusive averages. Note that these inclusive quantities can also be
presented in terms of event averages, namely
$\overline{x}=\langle X\rangle/\langle N\rangle$ and
$\overline{x^2}=\langle X_2 \rangle /\langle N\rangle$, where
$X_2\equiv \sum_{i=1}^Nx_i^2$.
The measure $\Phi$ was extended in 1999~\cite{Phi,Phi-Chem} for
multiplicity fluctuations.
For two
particle types, $A$ and $B$, the $\Phi$ measure was constructed by
setting $x_i=1$ if the $i$-th particle is of the $A$ type
and $x_i=0$ otherwise. One then finds
$\overline{x}=\overline{x^2}=\langle A\rangle/[\langle
A\rangle+\langle B\rangle]$ and
%$\overline{x^2}=\langle
%A\rangle/[\langle A\rangle+\langle B\rangle]$,
thus $\overline{x^2}-\overline{x}^2=\langle A\rangle \langle B
\rangle/[\langle A\rangle+\langle B\rangle]^2$ with $A$ and $B$
denoting particle numbers. Taking into account these
relations, and using $X=A$ and $N=A+B$ in Eq.~(\ref{phi_x}), the
expression for the $\Phi$ reads:
\begin{equation}\label{phi_n}
\Phi~=~\frac{\sqrt{\langle A \rangle \langle B \rangle}} {\langle
A+B\rangle }\left[ ~\left(\Sigma^{AB}\right)^{1/2}~-~1\right]~,
\end{equation}
where $\Sigma^{AB}$ is given by Eq.~(\ref{Psi}) with $C\equiv A+B
$.

A possible extension of $\Phi$ for the case of two motional
variables was not discussed up to now, however it can be
naturally done within the framework of the $\Sigma^{AB}$ and $\Delta^{AB}$
families presented here. It is also important to note that the
$\Phi$ measure extended to the study of the third moment preserves
its strongly intensive properties within the model of independent
sources~\cite{phi3}. Study of strongly intensive quantities which
include 3$^{rd}$ and higher moments of extensive quantities is
beyond the scope of this paper.

Another quantity frequently used to characterize the fluctuations
of particle numbers $A$ and $B$ was introduced in
2002~\cite{nudyn} as:
\eq{\label{nu}
%
%\sigma_{dyn}^2~\equiv~
\nu^{AB}_{dyn}~\equiv~\frac{\langle A(A-1)\rangle}{\langle
A\rangle^2}~+~\frac{\langle B(B-1)\rangle}{\langle
B\rangle^2}~-~2~\frac{\langle AB\rangle}{\langle A\rangle\langle
B\rangle}~.
}
Using Eq.~(\ref{Psi}) with $C\equiv  A+B $, one easily finds the
relation:
\eq{\label{nu-Psi}
\nu^{AB}_{dyn}~=~\frac{\langle A+B\rangle}{\langle A \rangle
\langle B\rangle }\left[~\Sigma^{AB}~-~1\right]~.
%\frac{\langle
%A~+~B\rangle}{\langle A\rangle \langle B\rangle }~.
%
}
Equation (\ref{nu-Psi}) shows that $\nu^{AB}_{dyn}$~, similar to
$\Sigma^{AB}$, is independent of fluctuations of the source
number, but it decreases as $\nu_{dyn}^{AB} \propto \langle
N_s\rangle ^{-1}$ and, thus, it is not an intensive quantity. Note
that the quantity $\langle C\rangle \nu_{dyn}^{AB}$, where $C$ can
be chosen as $A$, $B$, or $A+B$, is a strongly intensive quantity
from the $\Sigma$ family.
%It differs from $\Phi$ (Eq.~\ref{phi_n})
%only by the absence of the square root.
% the $\Sigma^{AB}$.
%
%
Despite the fact that specific examples of the $\Sigma^{AB}$
family were introduced and discussed a long time ago, the
$\Delta^{AB}$ family is proposed in this paper for the first time.

\section{Proof within the grand canonical ensemble}
Let us now prove within the grand canonical ensemble (GCE)
that the two families of quantities, $\Sigma^{AB}$ and
$\Delta^{AB}$, are strongly intensive. The proof  will be limited
to a case of particle multiplicities, i.e., $A$ and $B$ will be
number of particles of type A and B, respectively.
The GCE partition function $\Xi$ of the quantum gas which is a mixture
of different types of particles reads:
\eq{\label{Xigce}
 \Xi~=~\exp\left\{~V~\sum_j \eta_j~d_j
\int\frac{d^3p}{(2\pi)^3}~\ln\left[1~+\eta_j~
\lambda_j~\exp(-\epsilon_j/T)\right]\right\}~,
}
where $V$ and $T$ denote, respectively, the system volume and
temperature, $\lambda_j$  is the  fugacity which is related to
particle chemical potential $\mu_j$ as $\lambda_j\equiv
\exp(\mu_j/T)$, $d_j$ denotes the number of a particle internal
degrees of freedom, $\epsilon_j\equiv (m_j^2+{\bf p}^2)^{1/2}$ is
the particle energy with $m_j$ and ${\bf p}$ being its mass and
momentum,  $\eta_i=-1$ for bosons, $\eta_i=1$ for fermions, and
$\eta_i=0$ corresponds to the classical Boltzmann approximation.
The GCE averages are calculated as:
 \eq{\label{NA}
\overline{A}  \;&=\;
\frac{1}{\Xi}~\lambda_A\frac{\partial}{\partial
\lambda_A}~\Xi~=~V~\int\frac{d^3p}{(2\pi)^3}~\frac{d_A}
{\lambda_A^{-1}\exp(\epsilon_A/T)~+~\eta_A}
~\equiv~ V~n_{A}~,\\
 \overline{ A^2}\;&=\;
\frac{1}{\Xi}~\left(\lambda_A\frac{\partial}{\partial
 \lambda_A}\right)^2\Xi~=~V^2~n_A^2~+~
V~\int\frac{d^3p}{(2\pi)^3}~\frac{d_A\lambda_A^{-1}\exp(\epsilon_A/T)}
{\left[\lambda_A^{-1}\exp(\epsilon_A/T)~+~\eta_A\right]^2}~,
 \label{NA2}\\
 \overline{AB}
  \;&=\;
\frac{1}{\Xi}~\lambda_A\frac{\partial}{\partial
 \lambda_A}~\lambda_B\frac{ \partial}{\partial \lambda_B}~\Xi
~=~V^2~n_A n_B~,
 \label{NaNb-1}
 }
where $n_A=\overline{A}/V$ and $n_B=\overline{B}/V$ denote the
particle number densities. The corresponding expression for
$\overline{B}$ and $\overline{B^2}$ are obtained by replacing $A$
by $B$ in Eqs.~(\ref{NA}-\ref{NA2}).

For the GCE scaled variance one finds,
\eq{\label{omega-j}
\omega^*_A~\equiv~\frac{\overline{A^2}~-~\overline{A}^2}{\overline{A}}
%~\equiv~\frac{\langle N_j^2\rangle~-~\langle
%N_j\rangle^2}{\langle N_j\rangle }
~=~n_A^{-1}~\int\frac{d^3p}{(2\pi)^3}
\frac{d_A\lambda_A^{-1}\exp(\epsilon_A/T)}
{\left[\lambda_A^{-1}\exp(\epsilon_A/T)~+~\eta_A\right]^2}~.
}
It corresponds to the particle
number fluctuations at a fixed volume $V$. It is an intensive quantity
and depends only on $T$ and $\mu_A$. Note that $\omega_A^*>1$
for bosons, $\omega_A^*<1$ for fermions, and $\omega_A^*=1$ for
classical Boltzmann particles.
We introduce now volume fluctuations assuming that local
properties of the system within the GCE, i.e., the temperature and
chemical potentials, are volume independent. The volume fluctuations
will be described by the probability density function $F(V)$.
Thus, the full averaging  denoted as $\langle \ldots
\rangle $ will include both the GCE averaging
(\ref{NA}-\ref{NaNb-1}) at a fixed volume and an averaging over
the volume fluctuations:
 \eq{\label{NANA2}
\langle A\rangle   \;&=\; \langle V\rangle ~n_{A}~,~~~~
 \langle A^2\rangle \;=\;
%\frac{1}{\Xi}~\left(\lambda_A\frac{\partial}{\partial
% \lambda_A}\right)^2\Xi~=~
\langle V^2\rangle ~n_A^2~+~ \langle V\rangle~n_A~\omega_A^*~,
%~\int\frac{d^3p}{(2\pi)^3}~\frac{d_A\lambda_A^{-1}\exp(\epsilon_A/T)}
%{\left[\lambda_A^{-1}\exp(\epsilon_A/T)~+~\eta_A\right]^2}~,
% \label{NA2}
%\\
%
~~~~\langle AB\rangle
  \;=\;
%\frac{1}{\Xi}~\lambda_A\frac{\partial}{\partial
% \lambda_A}~\lambda_B\frac{ \partial}{\partial \lambda_B}~\Xi
\langle V^2\rangle ~n_A n_B~,
% \label{NaNb-2}
 %
 }
where %($k=1,2$)
%
%\eq{\label{Vk}
%
$\langle V^k\rangle \equiv\int dV~V^k F(V)$~ for $k=1,2$.
One finds,
\eq{\label{omegaA}
\omega_A ~\equiv~\frac{\langle A^2\rangle~-~\langle
A\rangle^2}{\langle A\rangle } ~=~\omega^*_A~+~n_A~\frac{\langle
V^2\rangle~-~\langle V\rangle^2}{\langle V\rangle}
~\equiv~\omega_A^*~+~n_A~\omega_V~.
}
The corresponding expression for $\omega_B$ is obtained by
replacing $A$ by $B$ in Eq.~(\ref{omegaA}).
Equations~(\ref{omegaA}) and (\ref{omega-A}) have a similar
structure. Namely, the first terms  $\omega^*_A$ or $\omega_a^*$
correspond to the particle number fluctuations at a fixed volume
$V$ or fixed number of sources $N_s$, respectively. The second
terms correspond to the contribution from the volume fluctuations
in Eq.~(\ref{omegaA}) and the fluctuations of the number of
sources in Eq.~(\ref{omega-A}).

Calculating (\ref{Psi}-\ref{G})  according to Eq.~(\ref{NANA2})
with $C=A+B$ one gets:
\eq{
\label{SD-GCE}
\Sigma^{AB}~=~\frac{1}{n_A+n_B}~\Big[n_B~\omega^*_A
~+~n_A~\omega^*_B\Big]~,~~~~
%~-~2\rho^*_{AB}
%\left(n_An_B~\omega_A^*\omega^*_B\right)^{1/2}~\right]~,\\
%
%\label{DeltaGCE}
%
\Delta^{AB}~=~\frac{1}{n_A+n_B}~\Big[n_B~\omega_A^*~-~n_A~\omega_B^*\Big]~.
}
Equation (\ref{SD-GCE}) proves that $\Sigma^{AB}$ and
$\Delta^{AB}$ are {\it strongly intensive} quantities as they are
strictly independent of average volume $\langle V\rangle$ and its
fluctuations $\omega_V$.

Note that in the GCE there are no correlations between the number
of different particle species, i.e., $\langle AB\rangle -\langle
A\rangle \langle B\rangle=0$. In the applications to hadron
production in high energy collisions  stable particles are
detected, whereas the GCE system includes also short
lived resonances which finally decay into stable particles.
These resonance decays increase  multiplicities of stable
particles and thus change  numerical values of
$n_A$, $n_B$, $\omega_A^*$, and $\omega_B^*$. If decay
products of a resonance R decay include both A and B hadrons 
a correlation between them appears and it can be expressed as:
\eq{\label{res-corr}
\langle AB\rangle - \langle A\rangle \langle B \rangle = \sum_R
\langle R \rangle \Big[\langle AB\rangle_R -\langle
A\rangle_R\langle B\rangle_R\Big]~ \equiv~\sum_R \langle R
\rangle~\rho_{AB}^R~,
%\left[\sum_r b_r^Rn_{A,r}^Rn_{B,r}^R -
%\sum_rb_r^Rn_{A,r}^R \sum_rb_r^Rn_{B,r}^R\right]~.
%
}
where $\langle R \rangle$ is a mean multiplicity of R and 
$\langle \ldots \rangle_R$ means the averaging over the
its all  decay channels. The measure $\Sigma^{AB}$
%(\ref{SD-GCE})
will have then the form:
\eq{\label{Sigma-res}
\Sigma^{AB}~=~\frac{1}{n_A+n_B}~\Big[n_B~\omega^*_A
~+~n_A~\omega^*_B~ ~-~2\sum_R n_R~ \rho^R_{AB}\Big]~,
%\left(n_An_B~\omega_A^*\omega^*_B\right)^{1/2}~\right]
%
}
and thus it remains a strongly intensive quantity.
\section{
Properties of ${\bf \Sigma^{AB}}$ and ${\bf \Delta^{AB}}$  within
Canonical and Micro-Canonical Ensembles}
For a large volume system in equilibrium  the particle number
distribution $P(A,B;V)$  in the GCE, canonical ensemble (CE), and
micro-canonical ensemble (MCE) can be written in a general form of
the bi-variate normal distribution (see Refs.~\cite{clt,BGG}):
 \eq{\label{P-bivar}
 & P(A,B;V)
 \;=\; \frac{1}{2\pi}\left[\omega^*_A~\omega^*_B\left(1-\rho_{AB}^{*2}\right)~
 \overline{A}~\overline{B}\right]^{-1/2}\;
         \nonumber \\
 & \times~ \exp\left[\; -\;\frac{1}{2\,(1-\rho_{AB}^{*2})}
 \left(\frac{(A-\overline{A})^2}{\omega^*_A~\overline{A}}
 \;-\; 2\,\rho^*_{AB}\;\frac{(A-\overline{A})(B-\overline{B})}
        {[\omega^*_A~\omega^*_B~ \overline{A}~\overline{B}]^{1/2}}
 \;+\; \frac{(B-\overline{B})^2}{\omega^*_B~\overline{B}} \right)\right]\;,
 }
where
%$V$ stands for the system volume, and
$\overline{A} \equiv n_AV~$ and $~\overline{B}\equiv n_B V$~ are
mean particle numbers.
%, with $n_A$ and $n_B$ denoting  the
%particle number densities.
Averaging at a fixed volume $V$ is defined as ($k=1,2$):
\eq{\label{av-V}
\overline{A^k}\equiv \sum_{A,B}
A^kP(A,B;V)~,~~~~\overline{B^k}\equiv\sum_{A,B} B^kP(A,B;V)~,~~~~~
\overline{AB}\equiv\sum_{A,B} AB ~P(A,B;V)~.
}
The straightforward calculations of (\ref{av-V}) with the
distribution function (\ref{P-bivar}) give:
 \eq{\label{rhoab}
 \frac{\overline{ A^2}~-~\overline{
        A}^2}{\overline {A}}~=~\omega^*_A ~,\qquad
 \frac{\overline{ B^2}~-~\overline{
        B}^2}{\overline {B}}~=~\omega^*_B ~,
~~~~~
 \frac{\overline{A\,B}~-~
 \overline{A}~\overline{B}}
                    {[\omega^*_A\omega^*_B~\overline{A}~\overline{B}]^{1/2}}
                    ~=~\rho^*_{AB}~.
}
Equation~(\ref{rhoab}) reveals the meaning of the parameters in
the distribution (\ref{P-bivar}) -- the scaled variances
$\omega^*_A$ and $\omega^*_B$, and the correlation coefficient
$\rho^*_{AB}$. In a mixture of relativistic ideal gases 
particle numbers are not conserved, and thus $A$ and $B$ fluctuate
in all statistical ensembles. This leads to non-zero positive
values of $\omega_A^*$ and $\omega_B^*$ which are approaching
constant values with system volume increasing to
infinity\footnote{ The volume dependence may be different for
system at the phase transition. For example, in a case of the
Bose-Einstein condensation one gets, $\omega \propto V^{1/3}$ at
$T=T_C$ and $\omega \propto V$ at $T<T_C$, as shown in
Ref.~\cite{BEC}.}. Particle correlations, and thus non-zero
$\rho^*_{AB}$, result from exact material and motional
conservation laws~\cite{CE}. Thus the correlation coefficient
$\rho^*_{AB}$ equals to zero in the GCE, and is non-zero in the CE
and MCE. The exact conservation laws also influence the values of
$\omega_A^*$ and $\omega_B^*$.
The  quantities like the particle number densities do not depend
on the choice of the statistical ensemble for large systems. This
means thermodynamical equivalence of the statistical ensembles. 
Below we
present results for the CE and MCE in the large volume limit 
in which all three statistical ensembles become
thermodynamically equivalent. Let us however stress that the
thermodynamical limits of the quantities (\ref{rhoab}) are 
different (see Ref.~\cite{CE} for details) in the GCE, CE, and MCE
ensembles.

We introduce now the volume fluctuations assuming that local
properties of the system (e.g., temperature and conserved charge
densities in the CE, or energy density and conserved charge
densities in the MCE) are volume independent for large enough
volumes. In this case, the distribution (\ref{P-bivar}) depends on
the system volume only through the average multiplicities. The
full averaging reads:
\eq{
\langle \ldots \rangle~ \equiv~\int dV F(V)~\sum_{A,B}
\ldots~P(A,B;V)~. \label{av-V2}
}
Calculating (\ref{Psi}-\ref{G})  according to Eq.~(\ref{av-V2})
with $C=A+B$ one gets:
\eq{
\label{Psia}
\Sigma^{AB}~&=~\frac{1}{n_A+n_B} ~\left[~n_B~\omega^*_A
~+~n_A~\omega^*_B ~-~2\rho^*_{AB}
\left(n_An_B~\omega_A^*\omega^*_B\right)^{1/2}~\right]~,\\
\label{Ga}
\Delta^{AB}~&=~\frac{1}{n_A+n_B}
~\left[~n_B~\omega_A^*~-~n_A~\omega_B^*~\right]~.
}
Equations (\ref{Psia}) and (\ref{Ga}) show that $\Sigma^{AB}$ and
$\Delta^{AB}$ are independent of the average system volume and its
fluctuations. For the CE and MCE this is valid if the 
volume fluctuates in a range in which all
three statistical ensembles are thermodynamically equivalent.

\vspace{0.2cm}
A unique determination of five intensive quantities, $n_A$, $n_B$,
$\omega^*_A$, $\omega^*_B$, and $\rho^*_{AB}$, from measurements
of $\langle A\rangle$, $\langle B\rangle$, $\langle A^2\rangle$,
$\langle B^2\rangle$, and $\langle AB\rangle$, is impossible as
$\langle V\rangle$ and $\omega_V$ are in general unknown. The
average particle multiplicities are given by $\langle
A\rangle=n_A\langle V\rangle$ and $\langle B\rangle=n_B\langle
V\rangle$. Therefore, only the ratio of particle number densities,
$r_{AB}\equiv \langle A\rangle/ \langle B\rangle=n_A/n_B$, can be
found from the measurements of $\langle A\rangle$ and $\langle
B\rangle$. The three strongly intensive quantities, $r_{AB}$,
$\Sigma^{AB}$, and $\Delta^{AB}$, allow a unique determination
of $\omega_A^*$ and $\omega^*_B$ in the GCE, if resonance decay
effects are absent. In this case there are no correlation between
$A$ and $B$ at fixed volume, and from Eq.~(\ref{SD-GCE})
one finds:
\eq{\label{omega-ab}
\omega^*_A~=~2\left(1+r_{AB}\right)\left[\Sigma^{AB}
~+~\Delta^{AB}\right]~,~~~~~
\omega^*_B~=~2\left(1+r_{AB}^{-1}\right)\left[\Sigma^{AB}
~-~\Delta^{AB}\right]~.
}
However, if $\rho^*_{AB}\neq 0$, as in the CE and MCE, or
including correlations due to resonance decays in the GCE, even the 
knowledge of all strongly intensive quantities is not sufficient to
reconstruct $\omega_A^*$, $\omega^*_B$, and $\rho_{AB}^*$ in a
unique way.

\section{Summary}
 In summary, in this paper we consider two families
of strongly intensive quantities $\Sigma^{AB}$ and $\Delta^{AB}$
which characterize  fluctuations of system properties.
While specific measures from the $\Sigma^{AB}$ family were
introduced already  in 1992~\cite{GM}, the $\Delta^{AB}$ family is
proposed in this paper for the first time. We prove within the
grand canonical ensemble
that both $\Sigma^{AB}$  and
$\Delta^{AB}$ quantities are strictly independent of volume and
volume fluctuations.
In the canonical and micro-canonical ensembles they are approximately
independent of volume and volume fluctuations for large enough systems.
Note that the $\Phi$ and $\nu_{dyn}^{AB}$ measures, which can be
expressed in terms of $\Sigma^{AB}$, have already been used
successfully to study transverse momentum  and particle ratio
fluctuations (see, e.g., Refs.~\cite{phi,Tim} and references
therein).  We hope that the results presented in this paper will
be useful in further analysis of fluctuations of hadron production
properties in collisions of relativistic particles.

\vspace{0.5cm} {\bf Acknowledgments.}~~ We thank V.V.~Begun,
W.~Greiner, M.~Hauer,  S. Mr\'owczy\'nski, and P.~Seyboth for
fruitful discussions.  This work was in part supported by the
Program of Fundamental Research of the Department of Physics and
Astronomy of NAS, Ukraine and the German Research Foundation
(grant GA 1480/2-1).

\end{document}